# Electron-hole coupling in high-$T_c$ cuprate superconductors

Alexander Brinkman[1,2] and Hans Hilgenkamp[1]

[1]*Faculty of Science and Technology and MESA+ Research Institute,*
*University of Twente, P.O. Box 217, 7500 AE Enschede, The Netherlands*

[2]*Département de Physique de la Matière Condensée,*
*University of Geneva, Quai Ernest Ansermet 24, 1211 Genève, Switzerland*

**From recent Hall effect measurements and angle-resolved photo-emission spectroscopy the interesting picture emerges of co-existing hole- and electron-like quasiparticle bands, both in electron- and hole-doped superconducting cuprates. We reflect on the idea that bosonic electron-hole pairs may be formed in the cuprates and on the possibility that these pairs undergo Bose-Einstein condensation. The relevance to high-$T_c$ superconductivity in the cuprates will be discussed.**

High critical temperature ($T_c$) superconductors are commonly categorized into two groups: hole-doped (p-type) and electron-doped (n-type) cuprates. To the former category belong, for example, the first high-$T_c$ material $La_{2-x}Ba_xCuO_4$ [1], $YBa_2Cu_3O_7$ [2], $Bi_2Sr_2Ca_{n-1}Cu_nO_{2n+4}$ [3] and the superconductor with the highest critical temperature $HgBa_2Ca_{n-1}Cu_nO_{2n+2}$ [4, 5]. The latter category is formed by the materials $Ln_{2-x}Ce_xCuO_4$, with Ln = La, Nd, Pr, Eu, or Sm [6]. In line with the classical BCS concepts, the superconductivity in these materials is considered to result from pairing of two holes into Cooperpairs with a charge of 2e for the p-type compounds and of two electrons into pairs of -2e charge for the n-type materials. The formation of bosonic pairs of two coupled fermionic particles has indeed been evidenced in the high-$T_c$ cuprates from the quantization of magnetic flux in a superconducting ring in units of the flux quantum $\Phi_0 =$ h/2e [7]. The mechanism of the pair formation in the high-$T_c$ cuprates is however still elusive and one of the unresolved questions is whether the mechanism depends principally on the sign of the charges constituting the pairs.

It is of interest to note that several experiments provide indications for the simultaneous presence of mobile electron-like quasiparticles and hole-like quasiparticles in various high-$T_c$ compounds, as in a two-band scenario. Already in 1988, Eagles proposed [8] that the anomalous temperature-dependence of the Hall coefficient, $R_H$, above $T_c$ for $YBa_2Cu_3O_7$ may be resulting from the combination of electron- and hole-conduction, each with their own carrier densities and mobilities. More recently, Hall coefficients have systematically been measured for cuprates throughout the high-$T_c$ cuprate phase diagram (see Fig. 1) as a function of temperature and doping. The emerging picture for hole-doped cuprates is a decrease, and at high temperatures a possible sign change, in $R_H$ for going from under- to overdoping [9, 10]. A saturation temperature can be distinguished for which $R_H$ reaches values close to the ones observed in the high-temperature limit. This saturation temperature decreases upon overdoping, approaching temperatures of the order of $T_c$ [9]. Although the mobilities of the charge carriers can vary and anomalous effects may play a role, the main character of the charge carriers appears to change gradually from hole-like to electron-like upon doping, see Fig. 1b.

Dagan *et al*. [11] have recently revealed an analogous dependence for the electron-doped compounds. In this case, the negative Hall coefficient increases to zero as a function of doping and crosses zero for overdoping, implying that the charge character changes from electron-like to hole-like. The Hall coefficient trends have schematically been depicted in Fig. 1b.

The sign change of the Hall-coefficient suggests that in parts of the cuprate phase diagram electron-like and hole-like quasiparticles co-exist, in different bands or at different positions on the Fermi surface. A further support for the notion that electrons and holes can co-exist in cuprate superconductors comes from Angle Resolved Photo-Emission Spectroscopy (ARPES). In ARPES measurements, the evolution of the Fermi surface topology has been investigated as a function of doping for different types of cuprates. Ino *et al*. [12] concluded for the hole-doped cuprate $La_{2-x}Sr_xCuO_4$ that the hole-type surface around the $(\pi,\pi)$ point of the Brillouin zone develops into an electron-type surface around $(0,0)$ upon doping. The hole-surface boundary in the nodal direction of underdoped cuprates was found to exist of an arc. This arc was found to gain intensity as function of doping in the case of $La_{2-x}Sr_xCuO_4$ [13] and as function of temperature in the case of $Bi_2Sr_2CaCu_2O_{8+\delta}$ [14]. The filling of the hole-band can result in the formation of an electron-band which is also gradually filled. The existence of these multiple bands was recently used by Ando *et al*. [10] to clarify the Hall effect in underdoped $La_{2-x}Sr_xCuO_4$.

In the electron doped cuprates ARPES measurements [15] revealed a Fermi surface with electron pockets at $(\pi,0)$ and $(0,\pi)$ that is gradually transformed into the hole surface around $(\pi,\pi)$ that is expected from local density approximation calculations.

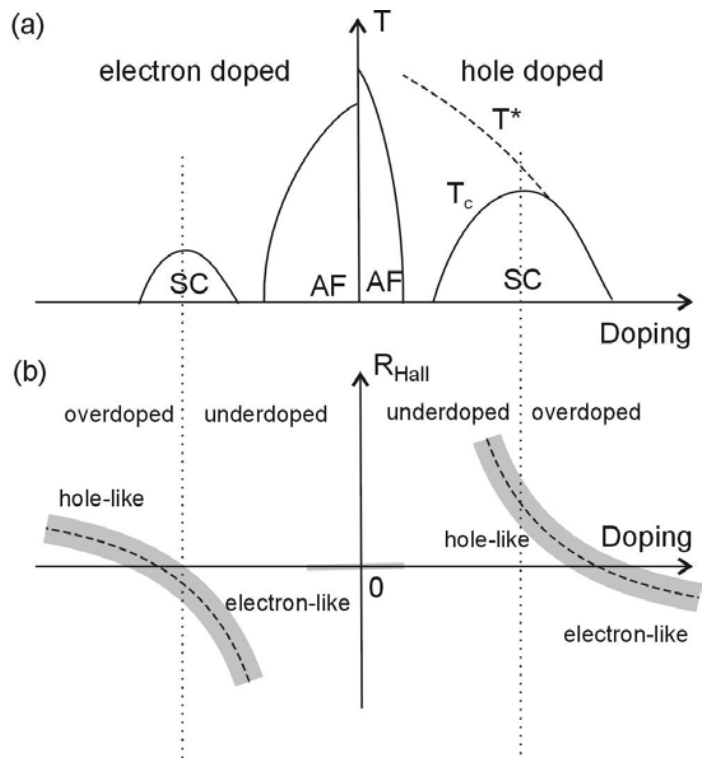

**Figure 1:** (a) Representation of the temperature versus doping phase diagram of high-$T_c$ cuprate superconductors, indicating regions of anti-ferromagnetism (AF) and superconductivity (SC). (b) Sketch of the Hall coefficient versus doping, at high temperatures.

One may wonder what the possible consequences could be of the attractive Coulomb interaction between the electron-like and hole-like quasiparticles when they are present simultaneously in these materials. Famous examples are excitons and electron-hole liquids [16]. Cooperpairs in BCS superconductors are composed of particles originating from the two opposite sides of the Fermi surface, resulting in a center of mass momentum much smaller than the Fermi-momentum $k_F$. Electrons and holes in general do not have opposite momenta and when a bosonic electron-hole pair is formed, the center of mass momentum can consequently be large, of the order of $2k_F$. As a result of this large momentum, condensation of the pairs cannot generally be described within the standard BCS model. In the following, we will reflect on the possibility that these electron-hole pairs undergo a Bose-Einstein condensation and whether or not this could relate to the high critical temperatures of the cuprate compounds.

A Bose-Einstein condensation (BEC) of atoms or molecules normally requires an ultra-low temperature. The condensate can form when the wave-packets of the individual bosons start to overlap, i.e. when the De Broglie wavelength $\lambda_{DB}$ approaches the mean distance between the bosons. For a gas of particles with mass $m$ and a thermal kinetic energy distribution, $\lambda_{DB}$ is given by [17]:

$$\lambda_{DB} = \sqrt{(2\pi \hbar^2 / m k_B T)}.$$

Now, let us consider a possible Bose-Einstein condensation of electron-hole pairs in the high-$T_c$ cuprates. The effective mass $m_{eh}$ of such a pair is not known, but its scale is set by the individual effective electron- and hole-masses, which are several orders of magnitude below the masses of atomic particles. As a calculation exercise, taking $m_{eh}$ to be $m_{eh} = 10\ m_e$, with the electron mass $m_e = 9.1 \times 10^{-31}$ kg, a De Broglie wave-length of about 7 nm is obtained for $T = 10$ K and about 2 nm for $T = 100$ K.

To estimate an upper bound for the density of electron-hole pairs we consider the density of the minority charge carriers. Interestingly, Eagles proposed that the majority charge carriers in optimally doped $YBa_2Cu_3O_7$ are in fact electrons, but due to their low mobility compared to the holes the sign of the Hall effect is positive [8]. For this hole density a value of $n = 5 \times 10^{20}$ cm$^{-3}$ was reported. This would imply a mean distance between the holes of 1-2 nm.

Although one may argue about the exact numbers, from this very coarse estimation it can be concluded that there are no unphysical parameter-values needed to arrive to the situation that the De Broglie wavelength is of the order of the mean distance between bosonic electron-hole pairs, if present, in high-$T_c$ cuprates at temperatures of the order of the critical temperature. We note also that this length-scale corresponds to the typical coherence lengths in the ab-plane of the cuprates. For the sake of further argumentation, let us therefore presume that a Bose-Einstein condensation of electron-hole pairs could occur at a certain temperature $T$. Would such a condensate necessarily be insulating or could it be (super)conducting?

When the charge of the electron-hole boson would be zero, one would intuitively say that the boson would not couple to the vector potential and that the condensate would be insulating. It is worth mentioning that in this case also interesting phenomena might arise, such as charge- and spin-density waves. However, as the electrons and holes constituting the pairs are perceived to have different ***k***-vectors, the pair states in real space are only short lived. Therefore the electron-hole pairs should not be viewed as electrically neutral entities traveling through the crystals, as for example in the usual picture of excitons in semiconductors. Rather, the electrons could be considered as hopping from hole to hole, with the electron- and hole-clouds moving on the average in different directions. From such a picture the possibility for (super)current transport follows naturally. We note that this scenario differs essentially from an exciton-mediated pairing of two charges with equal sign [18, 19], a modification of BCS theory in terms of a particle-hole channel [20], or the substantial work that has been performed on electron-proton coupling in metallic hydrogen [21]

The possibility of current flow due to the combined electron and hole transport can also be viewed differently. The concept of a holelike quasiparticle originates from the situation in which an electron is removed from the top of an otherwise full semiconductor valence band. The effective mass of the removed electron is inversely proportional to $d^2E(k)/dk^2$, which typically becomes negative at the top of the valence band. The remaining hole is commonly defined as having a positive mass and charge, to avoid complications of describing transport in terms of negative masses. The semiconductor-like picture of a single hole in an otherwise filled valence band is undoubtedly too limited to describe the situation in the high-$T_c$ cuprates. Extending it one may envisage the scenario that the Fermi level crosses the top of a band, leaving a small portion unfilled, which has a hole-like transport character. In that case it may be instructive to depart from a description in terms of holes and only consider the electron-electron interactions. For example, in curved Fermi surfaces, or in situations in which the Fermi surface contains multiple sheets, one could imagine that parts of the Fermi surface exhibit a negative curvature whereas other parts are positively curved. With this, the electrons that are available for current transport all have an equal negative charge, but their masses have unequal sign. An electron with negative mass could bind with an electron with positive mass thanks to their repulsive Coulomb interaction, forming a pair with charge -2e.

Closely related to the question in which way the electron-hole pairs couple to the vector potential, *A*, is the question whether or not electron-hole pairing could provide the necessary fluxoid quantization in units of $\Phi_0 = h/2e$. In the usual derivation of fluxoid quantization, the total Cooper pair momentum is ***P*** = 2m***v*** + e∗***A***, where *m* is the effective individual particle mass with velocity ***v*** and e* is the Cooper pair charge. Taking the contour integral of the second term over a closed path within a superconductor provides the total amount of enclosed flux, which in the standard case with |e*| = 2e can be shown to equal an integer number of flux quanta $\Phi_0$. The flux quantization condition thus couples the phase winding of the macroscopic wave function around a closed superconducting contour to the enclosed magnetic flux. As mentioned above, we presume that the electrons and holes can travel on the average in different directions and thus sustain a current. As a *Gedanken-experiment* let us consider that the holes travel clockwise and the electrons counter-wise around the ring. In a band picture, the motion of an electron with a given momentum ***k*** is equivalent to the motion of a hole with opposite

momentum –*k*. To derive the phase winding the electron-hole pair can thus be treated as if it was a pair of equal charges moving on the average in the same direction around the ring. Therefore the sum of the absolute values of the constituting charges is the relevant parameter, which is 2e, in line with the observed flux quantization.

In the above we have tried to paint a qualitative picture on possible consequences of the formation of bosonic electron-hole pairs in high temperature superconductors. As will be clear, there are many unresolved questions remaining, which need to be worked out also quantitatively. The most important one is arguably the question whether the $d_{x2-y2}$ order parameter symmetry [22] can be understood on the basis of such electron-hole pairing, which will need a further analysis on the spin-state of the electron-hole pairs. However, we propose the concept at this stage as we feel that it is worth a further consideration and the investigation of whether it can be ruled out as a contributing mechanism to high-$T_c$ superconductivity, or not. If high-$T_c$ superconductivity indeed relies on the Bose-Einstein condensation of electron-hole pairs a possible route towards higher critical temperatures would be to increase the minority charge carrier density and/or a reduction of the effective masses of the charge carriers.

The authors thank J.F. Annett, N.P. Armitage, N.W. Ashcroft, A.A. Golubov, A.B. Kuzmenko, J. Mannhart, D. van der Marel, H. Rogalla and C.C. Tsuei for valuable discussions. This work is supported by the Netherlands Organization for Scientific Research (NWO) and the Dutch Foundation for Research on Matter (FOM).

[1] J.G. Bednorz and K.A. Müller, *Z. Phys. B* **64**, 189 (1986).

[2] M.K. Wu, J.R. Ashburn, C.J. Torng, P.H. Hor, R.L. Meng, L. Gao, Z.J. Huang, Y.Q. Wang and C.W. Chu, *Phys. Rev. Lett.* **58**, 908 (1987).

[3] H. Maeda, Y. Tanaka, M. Fukutomi and T. Asano, *Jap. Jounal of Appl. Phys.* **27** – L209-L210 (1988).

[4] S.N. Putilin, E.V. Antipov, O. Chmaissem, M. Marezio, *Nature* **362**, 226-228 (1993).

[5] A. Schilling, M. Cantoni, J.D. Guo and H.R. Ott, *Nature* **363**, 56-58 (1993).

[6] Y. Tokura, H. Takagi and S. Uchida, *Nature* **337**, 345-347 (1989).

[7] C.E. Gough, M.S. Colclough, E.M. Forgan, R.G. Jordan, M. Keene, C.M. Muirhead, A.I.M. Rae, N. Thomas, J.S. Abell and S. Sutton, *Nature* **326**, 855 (1989).

[8] D.M. Eagles, *Solid State Commun.* **69**, 229 (1989).

[9] H.Y. Hwang, B. Batlogg, H. Takagi, H.L. Kao, J. Kwo, R.J. Cava, J.J. Krajewski, and W.F. Peck Jr., *Phys Rev. Lett.* **72**, 2636 (1994).

[10] Y. Ando, Y. Kurita, S. Komiya, S. Ono, and K. Segawa, *Phys. Rev. Lett.* **92**, 197001 (2004).

[11] Y. Dagan, M.M. Qazilbash, C.P. Hill, V.N. Kulkari, R.L. Greene, *Phys. Rev. Lett* **92** 167001 (2004).

[12] A. Ino, C. Kim, M. Nakamura, T. Yoshida, T. Mizokawa, A. Fujimori, Z.-X. Shen, T. Kakeshita, H. Eisaki, and S. Uchida, *Phys. Rev. B* **65**, 094504 (2002).

[13] T. Yoshida, X. J. Zhou, T. Sasagawa, W. L. Yang, P.V. Bogdanov, A. Lanzara, Z. Hussain, T. Mizokawa, A. Fujimori, H. Eisaki, Z.-X. Shen, T. Kakeshita, and S. Uchida, *Phys. Rev. Lett.* **91**, 027001 (2003).

[14] M.R. Norman, H. Ding, M. Randeria, J.C. Campuzano, T. Yokoya, T. Takeuchi, T. Takahashi, T. Mochiku, K. Kadowaki, P. Guptasarma, and D.G. Hinks, *Nature* **392**, 157 (1998).

[15] N.P. Armitage, F. Ronning, D.H. Lu, C. Kim, A. Damascelli, K.M. Shen, D.L. Feng, H. Eisaki, Z.-X. Shen, P.K. Mang, N. Kaneko, M. Greven, Y. Onose, Y. Taguchi, and Y. Tokura, *Phys. Rev. Lett.* **88**, 257001 (2002).

[16] W.F. Brinkman, T.M. Rice, P.W. Anderson and S.T. Chui, *Phys. Rev. Lett.* **28**, 961 (1972).

[17] W. Ketterle, *Rev. Mod. Phys.* **74**, 1131 (2002).

[18] W.A. Little, in *Novel Superconductivity*, S. Wolf and V. Kresin, Eds., p. 341 Plenum Press, New York, 1987.

[19] K.W. Wong and W.Y. Ching, *Physica C* **416**, 47 (2004).

[20] J.D. Fan and Y.M. Malozovsky, *Physica C* **364-365**, 101 (2001).

[21] K. Moulopoulos and N.W. Ashcroft, *Phys. Rev. Lett.* **66**, 2915 (1991).

[22] C.C. Tsuei and J.R. Kirtley, *Rev. Mod. Phys.* **72**, 969 (2000).